# Performance Evaluation of Distributed Computing Environments with Hadoop and Spark Frameworks

Vladyslav Taran*, Oleg Alienin, Sergii Stirenko, and Yuri Gordienko

National Technical University of Ukraine "Igor Sikorsky Kyiv Polytechnic Institute", Kyiv, Ukraine

*vladtkv@gmail.com

A. Rojbi,
CHArt Laboratory, (Human and Artificial Cognitions),
University of Paris 8,
2 Rue de la Liberté, 93526
Saint-Denis, France

*Abstract*— Recently, due to rapid development of information and communication technologies, the data are created and consumed in the avalanche way. Distributed computing create preconditions for analyzing and processing such Big Data by distributing the computations among a number of compute nodes. In this work, performance of distributed computing environments on the basis of Hadoop and Spark frameworks is estimated for real and virtual versions of clusters. As a test task, we chose the classic use case of word counting in texts of various sizes. It was found that the running times grow very fast with the dataset size and faster than a power function even. As to the real and virtual versions of cluster implementations, this tendency is the similar for both Hadoop and Spark frameworks. Moreover, speedup values decrease significantly with the growth of dataset size, especially for virtual version of cluster configuration. The problem of growing data generated by IoT and multimodal (visual, sound, tactile, neuro and brain-computing, muscle and eye tracking, etc.) interaction channels is presented. In the context of this problem, the current observations as to the running times and speedup on Hadoop and Spark frameworks in real and virtual cluster configurations can be very useful for the proper scaling-up and efficient job management, especially for machine learning and Deep Learning applications, where Big Data are widely present.

## I. Introduction

Recently due to the rapid development of social networks, mobile computing, Internet of Things (IoT), multimodal human-machine interactions, and other data-generating information and communication technologies (ICTs), the data are created and consumed in the avalanche way. These data now are actually Big Data, i.e. "data that's too big, too fast, or too hard for existing tools to process" [1]. The new opportunity to work with very large data sets (like ImageNet [2], YouTube8M [3], etc.) is one of the main recent abrupt development of the new machine learning (ML) approaches, especially Deep Learning [4], which recently emerged as the most promising trend in artificial intelligence (AI). According to IDC worldwide revenues for Big Data will grow from $130.1 billion in 2016 to more than $203 billion in 2020, at an annual growth rate about 12% [5]. The total volume of the data is estimated as 10 zettabytes in 2015, and "Big Data monetization" will become a major source of revenues from information-based products, as the world will create 180 zettabytes (1021 bytes) of data in 2025. As it is well-known, to perform analysis of any data, they first must be loaded into random access memory, and the fact, that the data volume is too big, does not allow processing them within a single computing node [6]. Distributed computing create preconditions for analyzing and processing big data by distributing the computations among a number of compute nodes. For the last years, many frameworks, that allow building clusters for distributed computations, were developed. For now, the most promising are Apache Hadoop [7] and Spark [8].

In this paper, we evaluate performance of distributed computing environments on the basis of Hadoop and Spark frameworks, particularly task running time, of Hadoop and Spark frameworks on our experimental cluster with Hadoop Distributed File System [7]. As a test task, we chose the classic use case of word counting in texts of various sizes. This paper is organized as follows. Section II provides the overview of Hadoop and Spark frameworks. Section III describes our experimental cluster settings and use case implementation of WordCount algorithm for Hadoop and Spark. Section IV presents results of our experiment, and Section V contains conclusions and future prospects for various data processing and recognition.

## II. Overview Of Frameworks

The core element of Hadoop framework is MapReduce [9] – the programming model for processing large data sets, distributed between cluster nodes. The idea of this model is splitting input data on part, computation of those parts on computing nodes – *NodeManager* and merging interim result into final solution on the main node – *ResourceManager*. For distributed data storing on cluster nodes, Hadoop has HDFS (Hadoop Distributed File System) [7]. The key elements of HDFS are *NameNode* – the main node that performs the functions of maintaining directories, files, and managing data blocks distributed between cluster nodes; and *DataNode* – the node that provides a physical storage space, process read/write requests from the main node. Advantages of Hadoop framework are that it doesn`t need specialized hardware for building computation cluster; has the lower cost of data storing and processing per terabyte; high scalability, productivity, fault-tolerance; and doesn`t need data conversion for storing them in distributed file system. Spark framework, compared with Hadoop, bring some improvements in distributed data processing and allows programs execution up to 100 times faster when processing data in memory, or 10 times – when processing data from disk [8]. Instead of MapReduce format, Spark can execute operators in Directed Acyclic Graph (DAG) format, which allows direct transferring interim result on the next processing stage without saving them in the distributed file system. In addition, Spark can perform in-memory processing. The popular *R* programming language supports a wide range of statistical methods, as well as machine learning operations such as linear and logical regressions, classifications and clustering [10]. To run programs written in *R*, Hadoop and Spark frameworks have the correspondent additional software like RHadoop [11] and SparkR [12].

## III. Experimental Environment

For performance evaluation of Hadoop and Spark, we deployed two clusters: the first cluster – the virtual one, which consists of three virtual computers and the second cluster – the real one with three servers. We setup Hadoop v2.7.3 and Spark v2.1.0 (in stand-alone mode) on our experimental hardware. As input, we used datasets of different size (from $10^5$ to $10^9$ bytes).

*A. Cluster Architecture*

The virtual version of cluster was deployed on Supermicro X8DTN server with Windows Server 2012 R2 operation system, 2 x Intel Xeon E5620 CPU–2.40Ghz and 32GB of RAM. For the virtual infrastructure, we used Microsoft Hyper-V (formerly known as Windows Server Virtualization), i.e. a native hypervisor, which can create virtual machines on x86-64 systems running Windows. For every virtual computer we allocated the following resources: CPU – 4 cores, 8GB of RAM, 128GB HDD storage. The real version of cluster was composed of three servers Supermicro X8DTN. One of them is designated as master, and the other two as slaves. The master node also performed the YARN worker role for execution of computational tasks along with cluster management by Apache Hadoop YARN (Yet Another Resource Negotiator). We use the operating system Ubuntu Server 14.04.5 LTS for every server. Each server has two Intel Xeon E5620 CPU –2.40Ghz, 32GB of memory and 1TB HDD storage.

*B. Cluster Installation*

The following stages were performed during clusters installation and configuration: 1) Setup Java v1.8.0_111; 2) Create dedicated cluster user; 3) Configure host file; 4) Configure SSH access between cluster nodes; 5) Setup Hadoop, configuration files modification (*core-site.xml, hdfs-site.xml, mapred-site.xml, yarn-site.xml*); 6) Setup Spark – Standalone mode; 7) Spark configuration file modification (*conf/slaves, conf/spark-defaults.conf*).

*C. Use Case Implementation*

To test performance of the mentioned frameworks we used WordCount, which is a simple application that counts the number of occurrences of each word in a given input set. For both Hadoop and Spark frameworks we used the classic WordCount algorithm and implemented it in Java.

## IV. Results and Discussion

WordCount tasks were run for data files of various sizes (from $10^5$ to $10^9$ bytes) on the pre-configured experimental clusters (see section III). Task running times were measured in seconds for Hadoop and Spark in relation to these dataset sizes. Their average ("Mean") and standard deviation ("Std") values were calculated on the basis of 10 trials for each run (Table 1), except for the biggest dataset in the virtual cluster configuration. For the better data visualization and comparison the running times for the datasets of various sizes were presented in the shape of double logarithmic plots (see Fig. 1). In general, on the basis of these results (Table 1, Fig. 1) the following observations were made:

- the real cluster configurations (the solid lines in Fig. 1) performs better than virtual ones (the dashed lines in Fig. 1) for both Hadoop and Spark frameworks;
- the Spark cluster configurations (the yellow lines in Fig. 1) performs better than Hadoop ones (the blue lines in Fig. 1) for both real and virtual versions;
- the standard deviations of the running times grow fast with the dataset sizes and nearly proportional to them;
- for the real cluster configurations the standard deviations of the running times for Hadoop are much lower than for Spark for the larger dataset sizes (626 MB and 5.5 GB), in contrast to the virtual cluster, where the standard deviations are not so different.

From the double logarithmic plots of these times vs. dataset sizes (Fig. 1) one can see that the running times grow very fast with the dataset size and faster than a power function even. This tendency is the similar for both Hadoop and Spark frameworks, and for both real and virtual version of cluster implementations. It means that scaling up the running tasks for the bigger datasets (=>100MB-1GB) could be very time demanding problem for both frameworks and real/virtual versions of clusters even for such relatively simple and standard tasks as WordCount task. The speedup was calculated (Table 1, Fig. 2) in comparison to the running time for the Hadoop framework on the real cluster version.

The following observations were made:

- in real cluster configuration the Hadoop framework (the blue solid line in Fig. 2) demonstrates the speedup values in relation to the virtual configuration (from 1.27 to 5.4) that are better than the Spark framework does (the blue dashed lines in Fig. 2);
- the Spark framework demonstrates the higher speedup values for the smaller datasets: steady speedup about

- 1.9-1.8 for sizes < 100MB, and about 1.1-1.3 for sizes =>100MB;
- the speedup values decrease with the growth of dataset sizes for both Spark and Hadoop frameworks.

To get maximum performance on our clusters we tried to determine experimentally the optimal configurations by varying the allocated memory up to 2048 MB for every Map and Reduce job.

TABLE I. THE RUNNING TIMES FOR DIFFERENT DATASET SIZES FOR WORDCOUNT TASKS: "MEAN" — MEAN VALUES; "STD" — STANDARD DEVIATIONS; "SPEEDUP" — THE CALCULATED SPEEDUP IN COMPARISON TO THE HADOOP FRAMEWORK ON THE REAL CLUSTER VERSION.

| Frame Work | Data Size | | | | | | | | | | | |
|---|---|---|---|---|---|---|---|---|---|---|---|---|
| | *287KB* | | | *8.7MB* | | | *626MB* | | | *5.5GB* | | |
| | *Mean* | *Std* | *Speedup* | *Mean* | *Std* | *Speedup* | *Mean* | *Std* | *Speedup* | *Mean* | *Std* | *Speedup* |
| *Hadoop* | 17.4 | 0.15 | 1.00 | 19.8 | 0.1 | 1.00 | 49 | 2 | 1.0 | 206 | 11 | 1.0 |
| *Spark* | 8.99 | 0.08 | 1.93 | 10.8 | 0.2 | 1.82 | 44 | 20 | 1.1 | 173 | 48 | 1.3 |
| | *Virtualization − Hyper-V* | | | | | | | | | | | |
| *Hadoop (Hyper-V)* | 22.01 | 0.11 | 0.79 | 24.55 | 0.03 | 0.8 | 99 | 6 | 0.50 | 1115 | - | 0.2 |
| *Spark (Hyper-V)* | 10.54 | 0.14 | 1.65 | 12.41 | 0.03 | 1.6 | 100 | 8 | 0.49 | 717 | - | 0.3 |

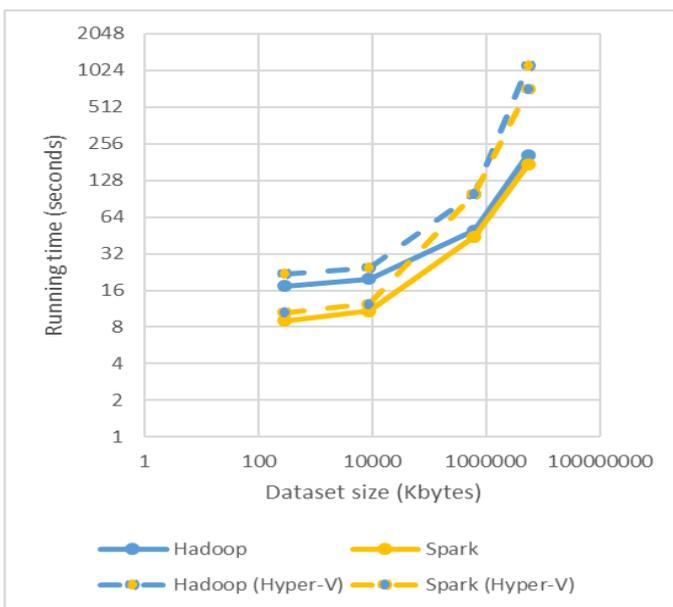

Fig 1. The averaged values of the running times vs. various dataset sizes.

We found that the further memory allocating increase does not give productivity improvements due to lack of memory to load all cores on cluster nodes. That is why the resource scheduler policy was changed to DominantResourceCalculator that takes into account the amount of available memory and cores on cluster.

## V. Conclusions and Future Work

As a result, in our experiment with the WordCount use case we found that Spark framework demonstrates the better performance in comparison with Hadoop. These results are generally confirmed by other similar researches and benchmarks for different other use cases [13-15]. In addition to this, we carried our experiments on 2 versions of cluster configurations (real and virtual) and found that the running times grow very fast with the dataset size and faster than a power function even (Fig. 1). As to the real and virtual versions of cluster implementations, this tendency is the similar for both Hadoop and Spark frameworks. Moreover, speedup decrease significantly with the growth of dataset size, especially for virtual version of cluster configuration.

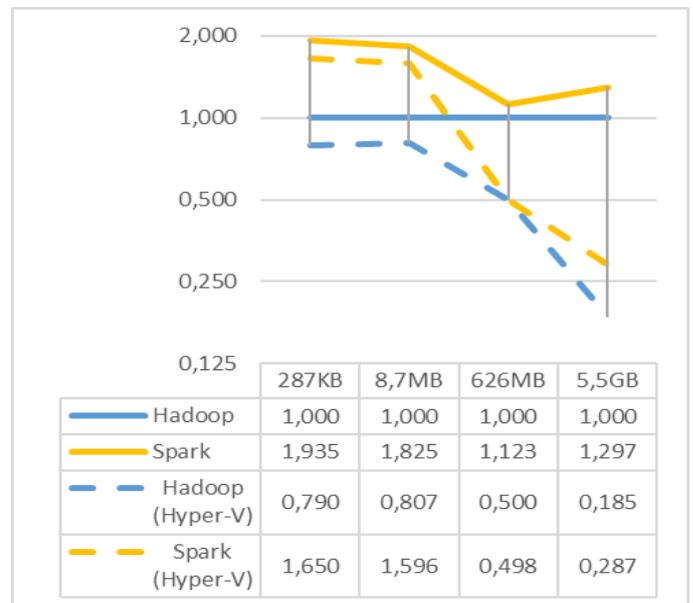

Fig 2. Speedup values for various cluster configurations in comparison to the running time for the Hadoop framework on the real cluster version.

In addition, it should be noted that Spark framework need less installation and configuration steps to achieve basic operability than during Hadoop installation and configuration process. Our recent comparative analysis of open source frameworks for machine learning with use case in single-threaded and multi-threaded modes on "MNIST data" proposed in 1998 to identify handwritten numbers was carried recently [16]. The performance tests for the de facto standard MNIST data set were carried out on the popular open source frameworks for machine learning (TensorFlow, Deep Learning4j, and H2O) designed for CPU and GPU platforms for single-threaded and multithreaded modes of operation. The high sensitivity of the running times to the dataset sizes was also observed and explained by the shortage of the memory on the available GPU-cards (12 GB for Tesla K40 with >30 GB of the swapped space on hard disk). In this context, the current observations as to the running time on Hadoop and Spark frameworks in real and cluster configurations can be very useful for the proper scaling-up and efficient job management in the context of ML and Deep Learning [4], where Big Data are widely present [2,3].

Despite the fact that we used the text-based datasets, the results obtained allow us to make the similar forecasts as to processing the data of various nature. Recently, Cloud-Fog-Dew computing paradigm was presented, which integrates several layers: Dew, Fog, and Cloud computing layer [17]. Usually, Dew computing layer envelopes the raw sensor data and basic multimodal actuator actions, which are concentrated, pre-processed, and resumed in the smallest scale local network (Dew) at the level of the IoT-controllers (individuals) and shared with the upper Fog computing layer. Fog computing layer includes the resumed IoT-controller data and advanced actuator actions, which are located in the medium scale regional network unit (Fog) at the level of the IoT-gateway and shared with the lower Dew computing layer and upper Cloud computing layer. Cloud computing layer gather the accumulated IoT-gateway data are thoroughly analyzed by ML methods to provide conclusions/decisions in the highest scale global network (Cloud) at the level of the global computing centers and delivered to the lower Fog and Dew Computing layer [18-19]. In the view of the rapidly growing number of IoT-enabled devices, the volume of data generated by them will increase hugely in the nearest years. That is why estimations of their processing times on various layers of Cloud-Fog-Dew hierarchy by Hadoop/Spark frameworks will be crucially important.

In addition to the available IoT data channels, which generate the abundant flow of data, the newly available multimodal (visual, sound, tactile, neuro and brain-computing, muscle and eye tracking, etc.) interaction channels produce the huge volume of multimedia data [20]. They concern the human-to-machine (H2M) and machine-to-human (M2H) interactions, including interactions with their environment (home, office, public places, etc.). The generated data can include 1D time series (digital signals like heartbeats, tremors, etc.), 2D figures (photo like X-ray images, video, etc.), 3D datasets (electroencephalography, computing tomography, etc.) related to individuals, groups of people, and nations even, for example, in the case of the health monitoring applications [19,20]. Again, taking into account the volume of data and necessity to process them in the shortest time, the efficient configuration of cluster (real or virtual) and frameworks could be crucial for the success of the Big Data processing. That is why performance analysis of these frameworks in the more complicated real scenarios, for example, on the very large data sets (like ImageNet [2], YouTube8M [3], etc.), which are in the great demand for artificial intelligence applications.

## Acknowledgment

The work was partially supported by Ukraine-France Collaboration Project (Programme PHC DNIPRO) (http://www.campusfrance.org/fr/dnipro) and Twinning Grant by EU IncoNet EaP project (http://www.inco-eap.net/).